# Electrically injected GeSn lasers on Si operating up to 100 K

Yiyin Zhou,[1,2] Yuanhao Miao,[1] Solomon Ojo,[1,2] Huong Tran,[1] Grey Abernathy,[1,2] Joshua M. Grant,[1,2] Sylvester Amoah,[1] Gregory Salamo,[3,4] Wei Du,[5] Jifeng Liu,[6] Joe Margetis,[7] John Tolle,[7] Yong-Hang Zhang,[7] Greg Sun,[8] Richard A. Soref,[8] Baohua Li,[9] and Shui-Qing Yu[1,4,*]

[1]Department of Electrical Engineering, University of Arkansas, Fayetteville, Arkansas 72701, USA
[2]Microelectronics-Photonics Program, University of Arkansas, Fayetteville, Arkansas 72701, USA
[3]Department of Physics, University of Arkansas, Fayetteville, Arkansas 72701, USA
[4]Institute for Nanoscience and Engineering, University of Arkansas, Fayetteville, Arkansas, 72701, USA
[5]Department of Electrical Engineering and Physics, Wilkes University, Wilkes-Barre, Pennsylvania 18766, USA
[6]Thayer School of Engineering, Dartmouth College, Hanover, New Hampshire 03755, USA
[7]School of Electrical, Energy and Computer Engineering, Arizona State University, Tempe, Arizona 85287, USA
[8]Department of Engineering, University of Massachusetts Boston, Boston, Massachusetts 02125, USA
[9]Arktonics, LLC, 1339 South Pinnacle Drive, Fayetteville, Arkansas 72701, USA
*Corresponding author: syu@uark.edu

**ABSTRACT**

The significant progress of GeSn material development has enabled a feasible solution to the long-desired monolithically integrated lasers on the Si platform. While there are many reports focused on optically pumped lasers, GeSn lasers through electrical injection have not been experimentally achieved yet. In this work, we report the first demonstration of electrically injected GeSn lasers on Si. A GeSn/SiGeSn heterostructure diode grown on a Si substrate was fabricated into ridge waveguide laser devices and tested under pulsed conditions. Special considerations were given for the structure design to ensure effective carrier confinement and optical confinement that lead to lasing. Lasing was observed at temperatures from 10 to 100 K with emission peaks at around 2300 nm. The minimum threshold of 598 A/cm$^2$ was recorded at 10 K and the threshold increased to 842 A/cm$^2$ at 77 K. The spectral linewidth of a single peak was measured as small as 0.13 nm (0.06 meV). The maximum characteristic temperature was extracted as 99 K over the temperature range of 10-77 K.

## 1. INTRODUCTION

Research advance in GeSn semiconductors has opened a new avenue for the development of Si-based optoelectronic devices [1-3]. With Sn content over 8%, GeSn turns into a direct bandgap material, which is essential for efficient light emission. Furthermore, the GeSn epitaxy is monolithic on Si and fully compatible with complementary metal-oxide-semiconductor (CMOS). The broad wavelength coverage also makes it versatile for mid-infrared applications, such as bio/chemical sensing, spectroscopy, and pyrometry [4,5]. All these advantages make GeSn material a promising candidate for the integrated light source on the Si platform that enables the system to be more compact, low-cost, efficient, and reliable.

In the last few years, there was considerable progress on the development of optically pumped GeSn lasers. The first GeSn laser was presented with 12.6% Sn composition operating at temperatures up to 90 K [6]. Later, higher Sn incorporation was reported to be beneficial on elevating the lasing temperature [7,8]. Further attempt on 20% Sn incorporation resulted in near room temperature lasing operation [9]. SiGeSn/GeSn heterostructure and multiple-quantum-well lasers were achieved with reduced threshold [10, 11] as well as elevated operating temperatures [12]. Efforts on strain engineering of the GeSn lasers showed great improvements of device performance as an alternative route to incorporating more Sn. Chrétien et. al showed the laser operating temperature as high as 273 K with 16% Sn composition [13]. A continuous wave optically pumped laser was reported with Sn composition as low as 5.4% in a tensile strained disk structure [14]. Thus far, all GeSn lasers were reported using optically pumping and how to achieve electrically injected lasers as predicted earlier [15, 16] remains elusive.

Here, we present the first demonstration of electrically injected GeSn diode lasers. The GeSn/SiGeSn double-heterostructure was grown which ensures the carrier-and-optical confinement. To address the hole leakage due to a type-II band alignment between GeSn and the top SiGeSn barrier, the p-type top SiGeSn layer was designed to facilitate the hole injection. The ridge waveguide GeSn lasers were fabricated and pulsed lasing was observed up to 100 K. The threshold was measured at 598 A/cm$^2$ at 10 K. The characteristic temperature $T_0$ was extracted from 76 to 99 K at temperature range of 10-77 K for different devices.

## 2. EXPERIMENT

The laser diode structure was grown via an industry standard chemical vapor deposition reactor using commercially available precursors of SiH$_4$, GeH$_4$, and SnCl$_4$ on a 200-mm (100) Si substrate. Five epitaxial layers were grown from bottom to top: i) a nominal 500-nm-thick strain-relaxed Ge buffer layer, with n-type doping of $1\times10^{19}$ cm-3; ii) a 700-nm-thick GeSn buffer layer using the spontaneous relaxation enhanced growth method [17], with nominal Sn composition from 8% (bottom) to 11% (top), and n-type doping of $1\times10^{19}$ cm$^{-3}$; iii) a nominally intrinsic 1000-nm-thick Ge$_{0.89}$Sn$_{0.11}$ active layer; iv) a 170-nm Si$_{0.03}$Ge$_{0.89}$Sn$_{0.08}$ cap layer with p-type doping of $1\times10^{18}$ cm$^{-3}$; and v) a 70-nm Si$_{0.03}$Ge$_{0.89}$Sn$_{0.08}$ Ohmic contact layer with p-type doping of $1\times10^{19}$ cm$^{-3}$. All doping growth was done in-situ by introducing corresponding doping gases. The cross-sectional schematic of the laser device is shown in Fig. 1(a). The compositions of Sn and the layer thickness were measured by x-ray diffraction and transmission electron microscopy techniques.

After the growth, the sample was fabricated into ridge waveguide laser structures and then cleaved into individual devices with the cavity lengths of 0.3, 0.5, 0.8, and 1.7 mm. The 80-µm-wide ridges were formed by wet etching. The etching depth was controlled at 1.4 µm to expose the GeSn buffer layer for metal contacts. Electron beam evaporated Cr and Au were deposited as the both p and n electrodes through a lift-off process with the thickness of 10 and 350 nm, respectively. The Si substrate was lapped down to 140 µm thick, followed by cleaving to form the Fabry-Perot cavity. Finally, the devices were wire-bonded to a Si chip carrier and mounted in a cryostat for low temperature measurements.

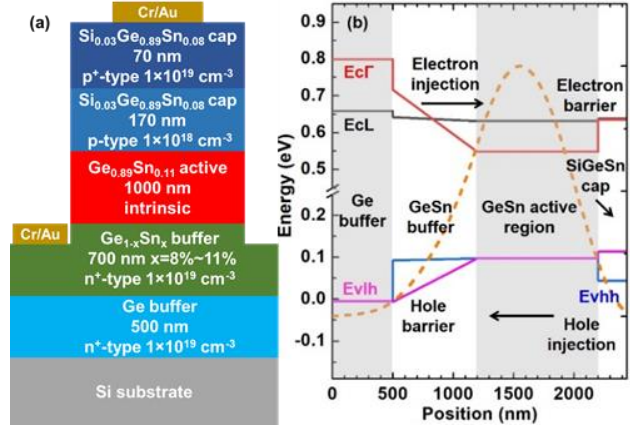

Fig. 1. (a) Cross-sectional schematic of laser device; (b) Calculations of band structure and profile of fundamental TE mode. Band structure shows type II alignment between GeSn active and SiGeSn cap layers at LH band. Mode profile shows 75% of the optical field overlapped with the GeSn active region.

The current-voltage (IV) measurement was performed using a direct current source measurement unit. For the pulsed measurement, a pulsed high compliance voltage source was used to drive the laser and the current was monitored by a calibrated magnetically coupled current meter. The repetition rate of 1 kHz and pulse width of 700 ns were used for the spectra and light output vs injection current (LI) measurements. The electroluminescence and photoluminescence (PL) emission signal was collected and analyzed through a monochromator (10 nm resolution) and liquid nitrogen cooled InSb detector (response range 1 - 5.5 μm). The high-resolution spectra were obtained by using a Fourier-transform infrared spectroscopy (FTIR) instrument equipped with a liquid nitrogen cooled InSb detector. Step-scan mode was used with a 0.25 cm$^{-1}$ resolution for the measurement.

## 3. RESULTS

The device band edge diagram at 300 K was calculated and plotted in Fig. 1(b). Four sub-bands including indirect L ($E_{cL}$) and direct Γ ($E_{cΓ}$) valleys in the conduction band (CB) and heavy hole ($E_{vhh}$) and light hole ($E_{vlh}$) in the valence band (VB) were considered. The following features are obtained from Fig. 1(b): i) $Ge_{0.89}Sn_{0.11}$ active layer has a direct bandgap with the energy difference of 84 meV between L and Γ valleys. The full relaxation of this layer results in the heavy hole (HH) and light hole (LH) band overlap; ii) In the CB, both Γ and L valleys feature type-I alignment due to the wider bandgap energies of $Si_{0.03}Ge_{0.89}Sn_{0.08}$ cap and GeSn buffer. Note that the Sn composition increases in the GeSn buffer (8%~11%) along the growth direction, leading to the decrease of both Γ and L valleys in energy with the Γ more rapidly than the L valley; iii) In the VB, the HH band features type-I band alignment. The LH exhibits type-II band alignment at the cap/active layer interface, due to the tensile strain within the $Si_{0.03}Ge_{0.89}Sn_{0.08}$ cap.

The fundamental TE mode was plotted (dashed curve) to show the optical field distribution. The refractive index for each layer was taken from our previous study [18]. The optical confinement factor (optical field confined in $Ge_{0.89}Sn_{0.11}$ active region) was calculated as 75% using the wavelength at 2.3 μm.

The typical pulsed LI curves from the 0.8-mm cavity length device were plotted in Fig. 2(a) at temperatures from 10 to 100 K (maximum lasing temperature). The threshold current densities are measured as 0.74 and 3.9 kA/cm$^2$ at 10 and 100 K, respectively. At 10 K, the emission shows a saturation feature at 7.5 kA/cm$^2$.

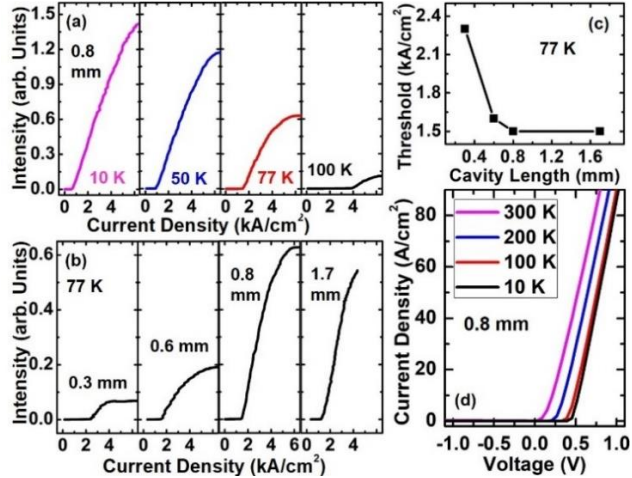

Fig. 2 (a) LI curves of the 0.8 mm cavity length device from 10 to 100 K; (b) LI curves at 77 K for four devices with different cavity lengths; (c) Threshold of each device at 77 K; (d) Temperature-dependent IV of the 0.8 mm cavity length device.

The LI characteristics of devices with different cavity lengths were studied, as shown in Fig. 2(b) at 77 K. The threshold current densities were measured as 2.3, 1.6, 1.5, and 1.5 kA/cm$^2$ for devices with cavity lengths of 0.3, 0.6, 0.8, and 1.7 mm, respectively, as shown in Fig. 2(c). As cavity length (L) increases from 0.3 to 0.6 mm, the dramatically decreased lasing threshold is mainly due to the reduced mirror loss ($\propto 1/L$). When the cavity length is longer than 0.6 mm, other loss mechanisms became dominant so that a very little change in threshold was obtained. Moreover, the saturated emission intensity increases as L increases, except for the 1.7 mm device (no higher current could be applied due to device damage).

The typical IV characteristics of the 0.8-mm cavity length device were measured at various temperatures, as plotted in Fig. 2(d). The series resistance is extracted as 2.85 Ω. From 300 to 10 K, the device turn-on voltage ranged from 0.03 to 0.12V. The small turn-on voltage was due to the narrow bandgap of GeSn. The increased turn-on voltage at the lower temperature is the result of increased bandgap energy.

The emission spectra below and above threshold were investigated. Figure 3(a) shows the spectra of the 0.3-mm cavity length device under various current injection levels at 10 K. Below the threshold, the peak full-width half max (FWHM) was 15.3 meV, while above the threshold the FWHM became ~5 meV, as shown in Fig. 3(a) inset. Note that the relative broad peak linewidth of 5 meV is due to the spectral resolution of 10 nm. At 10 K, the lasing emission peak was observed at 2250 nm.

Figure 3(b), (c), and (d) show the emission spectra of the 1.7-mm cavity length device at 10, 77, and 100 K, respectively. At each temperature, as the injection current density increases from below to above the threshold, the significantly increased peak intensity and reduced FWHM were observed, both being the evidence of the lasing characteristic. At 10 K, the measured peak position is the same as for the 0.3-mm device, i.e., at 2250 nm. At 100 K, the lasing peak redshifts to 2300 nm as expected due to the narrowed bandgap. The log-scale plot of Fig. 3(d) is shown in (e). Above the threshold, the stimulated emission peak stands out from the broad spontaneous emission.

The far-field pattern was measured at the cross-section plane 4 cm away from the laser facet. Two major peaks were observed, indicating multi-mode operation, as shown in Fig. 3(f). The FWHMs of the major peaks at the center are estimated around 12 and 16 degrees along the fast and slow axis, respectively.

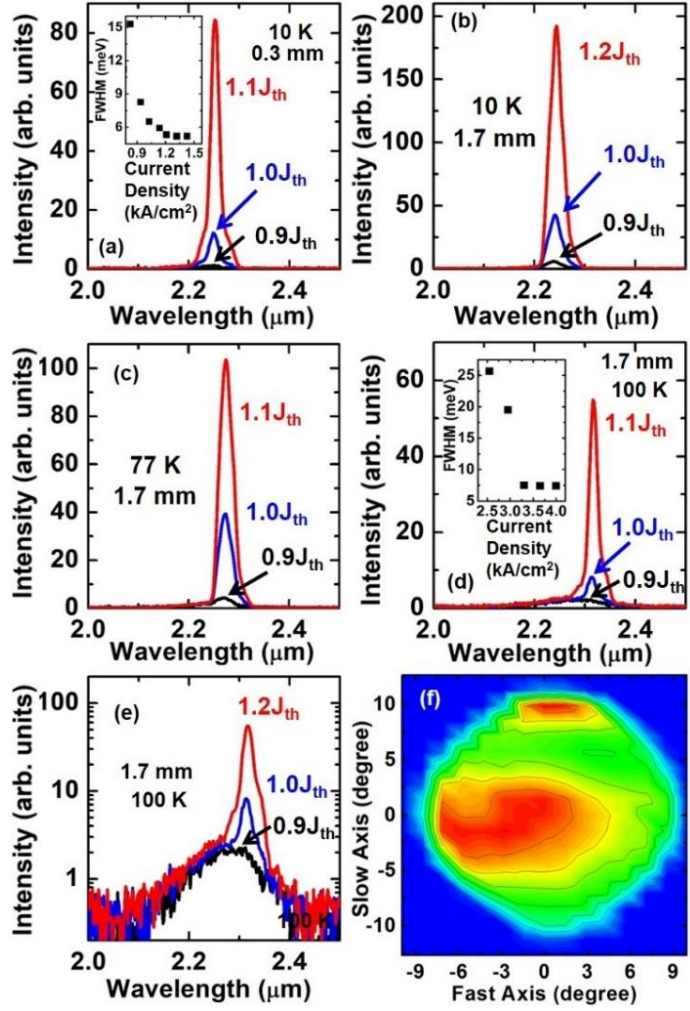

Fig. 3. Emission spectra at various current injection levels. (a) 0.3 mm device at 10 K. Inset: extracted FWHM vs current density (spectral resolution of 10 nm); (b), (c), and (d) 1.7 mm device at 10 K, 77 K, and 100 K. Inset in (d): extracted FWHM vs current density; (e) log-scale plot of spectra in (d); (f) far field pattern from 1.7 mm device at 77 K.

To further study the lasing characteristic, high-resolution spectra were measured using a FTIR. Figure 4 shows the spectra of the 0.8-mm cavity length device at 77 K under various current injections. Above the threshold, the multi-mode lasing characteristic was clearly observed. The minimum FWHM of the individual peak is measured as 0.13 nm (or 0.06 meV). The dramatically reduced peak linewidth under higher injection is one of the lasing characteristics. At above 1.42 × threshold, the peak at 2307 nm dominates the lasing spectrum.

The temperature-dependent threshold for each device was analyzed. The characteristic temperature $T_0$ was extracted using the empirical relation [19] $J_{th} = J_0 exp(T/T_0)$, where $J_{th}$ is the threshold current density, $J_0$ is a constant, T is the temperature. For each device, two regions, from 10 to 77 K and from 77 to 100 K can be clearly observed, as shown in Fig. 5 (except the 0.3-mm device whose maximum lasing temperature is 90 K). From 10 to 77 K, three devices (0.3, 0.6 and 0.8 mm) exhibited $T_0$ above 90 K, while the 1.7-mm device showed $T_0$ of 76 K. From 77 to 100 K, three devices show $T_0$ values close to ~30 K, as a result of the significantly increased threshold.

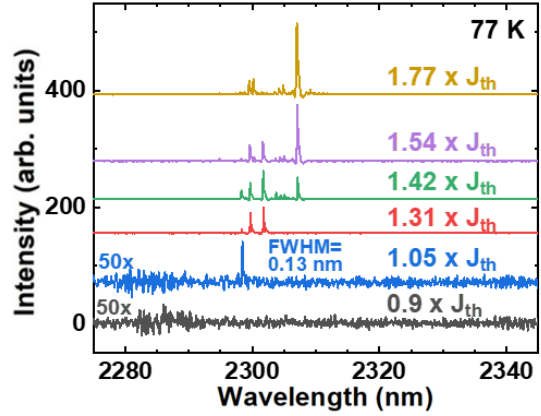

Fig. 4. High-resolution spectra of 0.8 mm cavity length device at 77 K under various current injections.

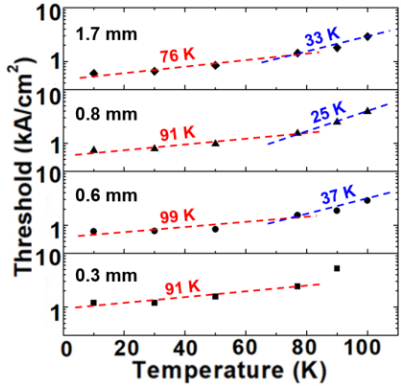

Fig. 5. Extracted $T_0$ for each device. For the 0.3 mm device, the data point of threshold at 90 K was excluded from data fitting.

## 4. DISCUSSION

In comparison with obtaining optically pumped lasers, there are additional considerations for both carrier confinement and optical confinement in this work for the device structure design. In considering the carrier confinement in these laser devices, the result of band structure calculation shown in Fig. 1(b) reveals that a type-II band alignment between the GeSn active and SiGeSn cap layers in LH at VB. The type-II alignment is originated from the tensile strain of the SiGeSn cap lifting the LH band above that in the GeSn active layer. As a result, the hole leakage at the top surface could lead to an increase in the lasing threshold. Therefore, this design is not ideal for optically pumped devices. However, for electrically injected devices, in order to address this poor hole-confinement issue, the top SiGeSn layer was intentionally doped as p-type in this work. As the holes are injected from the top SiGeSn cap layer, they are forced to flow to the GeSn active region. Since there are hole barriers in both HH and LH bands at VB between GeSn active and GeSn buffer layers, the holes could be confined in the active region to facilitate the population inversion. While the electrons are injected from the bottom GeSn buffer with the n-type contact, the electron barrier in Γ valley (lower band than L valley) in CB between the GeSn active and the top SiGeSn cap prevent the leakage of electrons and confine them in the active region. This p-i-n device structure design, rather than the n-i-p structure which may utilize the p-type unintentionally background doping of the GeSn buffer, effectively minimizes the hole leakage and enhances the carrier confinement.

In order to increase the optical confinement, it is necessary to address the small difference in refractive index between Ge (4.03 at 2.3 μm) and GeSn (4.1~4.2 with different Sn %). To increase the mode overlap with the GeSn active layer, an overall 240-nm-thick SiGeSn cap layer was grown on top of the GeSn active layer, which pushes the peak intensity of the optical field into the active region, resulting in a 75% mode overlap with the GeSn core layer whose thickness is 1000 nm, as shown in Fig. 1(a). For optically pumped laser devices, the optical field can be well confined since there is nothing but the air above the cap layer. However, for the electrically injected laser devices, due to the metal contact above the cap layer, the thickness of the SiGeSn cap needs to be carefully optimized to minimize the optical loss via the metal thin film. The current thickness is selected as a compromise of the SiGeSn growth capability and the metal optical loss.

The diode turn-on voltage is estimated as 0.06 V at 300 K after subtracting the potential drop at cables, which is smaller than the previous reported value of ~0.1 V with 6% Sn diodes [20]. This is expected as the 11% Sn in this work features narrower bandgap and higher intrinsic carrier concentration, which leads to smaller turn-on voltage as observed.

The lasing spectra in Fig. 3 were further examined with the PL study of the active region. The PL spectra were taken with an unprocessed sample (A) and with a cap-removed sample (B, where the SiGeSn cap was etched away and the GeSn active layer was exposed at the very top). For sample A, the PL peak at 2140 nm (0.579 eV) was observed at 10 K, which is close to the calculated transition energy from Γ valley to HH band (0.584 eV) in the SiGeSn cap. Considering the penetration depth (less than 100 nm) of the 532-nm excitation laser and 240-nm-thick SiGeSn cap, the major emission is from the cap layer. The smaller energy peak can be attributed to the tensile strain. For sample B, since the SiGeSn cap was completely removed, the emission is from the GeSn active layer. The PL peak at ~2400 nm was obtained at 10 K. Compared to the PL peak position, the lasing peak at 2250 nm features a clear blue-shift, which can be attributed to the typical band filling effect. In addition, the peak linewidth of ~0.06 meV at 77 K was extracted from lasing spectra, while the linewidths of the PL peaks are 32 and 41 meV for samples A and B, respectively. The dramatically reduced peak linewidth indicates the onset of lasing.

$T_0$ of ~90 K in the temperature range below 77 K is comparable with earlier reported III-V double heterostructure laser diodes [21, 22]. As temperature increases above 77 K, the carriers in the GeSn active region could gain sufficient thermal energy to overcome the barriers and leak in to the SiGeSn cap and GeSn buffer layers, resulting in significantly increased lasing threshold. This leads to the dramatically reduced $T_0$ above 77 K, ranging from 25 to 33 K, as shown in Fig. 5. By optimizing the design of cap and buffer layers, the improved carrier confinement at higher temperatures can be obtained and thus higher $T_0$ is expected.

In terms of further improving the device performance, investigations of new structure designs are underway, which include: (1) Increasing the Sn content to directly increase the possibility of electrons distributing in the Γ valley, therefore reducing the carriers leaking to indirect valley, (2) Adding an SiGeSn buffer on the n layer to provide enhanced hole confinement, (3) Optimizing the waveguide geometry to eliminate the higher order leaky modes, and (4) Reducing doping levels to minimize the free carrier absorption.

## 5. CONCLUSION

We have demonstrated electrically injected GeSn/SiGeSn heterostructure lasers that were grown on a Si wafer using a commercial CVD reactor. The lasing threshold of 587 A/cm$^2$ at 10 K was obtained. The multi-mode lasing characteristics were confirmed by high-resolution spectra. The maximum lasing temperature was measured as 100 K with 2300 nm peak wavelength. The p-i-n structure design enhances the carrier confinement by reducing the hole leakage through the type-II band aligned cap layer. The maximum $T_0$ of 99 K was obtained from a 0.6-mm cavity length laser device.

**FUNDINGS**

Air Force Office of Scientific Research (AFOSR) (FA9550-18-1-0045, FA9550-19-1-0341). Dr. Wei Du appreciates support from Provost's Research & Scholarship Fund at Wilkes University.